\documentclass[twocolumn%
              ,superscriptaddress%
              ,aps%
              ,prl]{revtex4}

\usepackage{amsmath,amssymb}
\usepackage{graphicx}

\begin{document}

\title{Plasma vs Drude modelling of the Casimir force: beyond the proximity force
approximation}

\author{Michael Hartmann}
\affiliation{Universit{\"a}t Augsburg, Institut f{\"u}r Physik, 86135 Augsburg, Germany}
\author{Gert-Ludwig Ingold}
\affiliation{Universit{\"a}t Augsburg, Institut f{\"u}r Physik, 86135 Augsburg, Germany}
\author{Paulo A. Maia Neto}
\affiliation{Instituto de F{\'i}sica, Universidade Federal do Rio de Janeiro, CP 68528,
Rio de Janeiro RJ 21941-909, Brazil}

\date{\today}

\begin{abstract}
We calculate the Casimir force and its gradient between a spherical and a planar
gold surface. Significant numerical improvements allow us to extend the range of
accessible parameters into the experimental regime. We compare our numerically
exact results with those obtained within the proximity force approximation (PFA)
employed in the analysis of all Casimir force experiments reported in the
literature so far. Special attention is paid to the difference between the Drude
model and the dissipationless plasma model at zero frequency. It is found that
the correction to PFA is too small to explain the discrepancy between the
experimental data and the PFA result based on the Drude model. However, it
turns out that for the plasma model, the corrections to PFA lie well outside the
experimental bound obtained by probing the variation of the force gradient with
the sphere radius [D.~E.\ Krause \textit{et al.}, Phys.\ Rev.\ Lett.\ \textbf{98},
050403 (2007)]. The corresponding corrections based on the Drude model are
significantly smaller but still in violation of the experimental bound for small
distances between plane and sphere.
\end{abstract}

\maketitle

The last decades have witnessed a surge in precise measurements of the Casimir
interaction \cite{Bordag2009,Klimchitskaya2009,Decca2011,Lamoreaux2011}.
Instead of the theoretical paradigm of two parallel metallic plates
\cite{Casimir1948}, most experiments adopt the plane-sphere geometry to avoid
misalignment. This geometry is in principle amenable to an exact description by
the scattering approach \cite{Lambrecht2006,Emig2007,Rahi2009}. The Casimir
interaction energy was calculated in the plane-sphere geometry at zero
temperature for perfect \cite{Maia2008,Emig2008} and real metals
\cite{Canaguier2009}, as well as for finite temperatures
\cite{Canaguier2010,Canaguier2010PRA} and in the high-temperature limit
\cite{Canaguier2012,Bimonte2012,Bimonte2017}. In spite of those recent
theoretical developments, the analysis of the plane-sphere experiments to this
date has relied exclusively on the heuristic proximity force approximation
(PFA), also known as Derjaguin approximation \cite{Israelachvili}.

\begin{figure}
 \begin{center}
  \includegraphics[width=0.8\columnwidth]{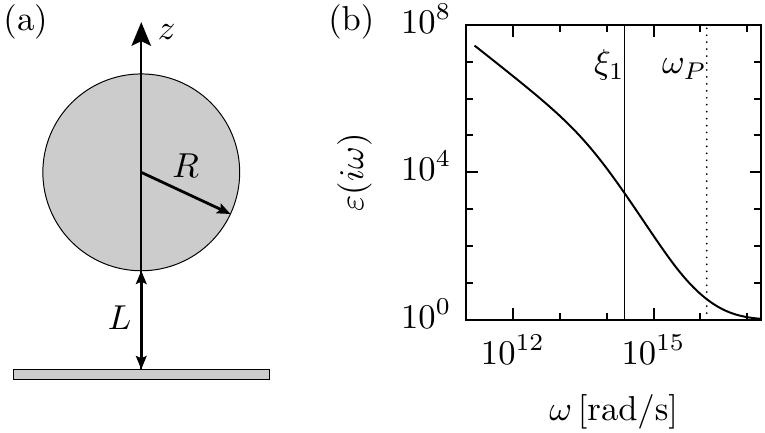}
 \end{center}
 \caption{(a) Sphere of radius $R$ separated from the plane by a distance $L$.
   (b) Frequency dependence of the permittivity of gold used in the numerical
   calculations. The vertical solid line indicates the first Matsubara frequency
   $\xi_1$ while the dotted line indicates the plasma frequency $\omega_P$.}
 \label{fig:geometry}
\end{figure}

Experimentally a measurable force signal requires the radius $R$ of the sphere
to be much larger than the separation $L$ between sphere and plane, cf.\
Fig.~\ref{fig:geometry}a. For probing distances in the micrometer range, coated
macroscopic lenses with radii of more than $10\,$cm leading to aspect ratios
$R/L\sim 10^5$ were used \cite{Lamoreaux1997,Sushkov2011}. In experiments
exploring the Casimir interaction in the sub-micrometer regime smaller aspect
ratios of $R/L \sim 10^3$ were realized \cite{Krause2007,Decca2007,Decca2007EPJC,Chang2012,Banishev2013,Bimonte2016}. However, up to now
these aspect ratios were out of reach for numerically exact computations.
Within the scattering approach, the required number of multipoles scales like
$R/L$. In practice, the number of multipoles so far was limited to $\ell \sim
500$ allowing for aspect ratios of $R/L \sim 100$ \cite{CanaguierThesis}. Such
calculations are capable of addressing recently proposed experiments based on
optical tweezers as a tool for probing femtonewton Casimir forces well outside
the validity of PFA \cite{Ether2015}, but are not suited for describing typical
Casimir force experiments.

In this letter, we significantly extend the range of numerically accessible
aspect ratios to values of $R/L \sim 10^3$ and report on results for the
Casimir force and force gradient. We take the parameters corresponding to the
experiments in Refs.~\onlinecite{Krause2007,Decca2007,Decca2007EPJC,Chang2012} with gold
surfaces at room temperature, but our approach also opens the way to calculate
exact results for a variety of recent experiments with similar aspect ratios
and different materials like magnetic materials \cite{Banishev2013} and layered
surfaces \cite{Bimonte2016}. The key ingredients allowing us to treat
experimentally relevant aspect ratios are a new symmetrical representation of
the round-trip scattering operator and a state-of-the-art algorithm for
evaluating determinants of hierarchical matrices.

Although PFA is expected to provide the correct leading divergence in the limit
$R/L\rightarrow \infty$, the magnitude of the correction to PFA under real
experimental conditions was not known until now.
In Ref.~\onlinecite{Krause2007}, the
force gradient variation with $R$ was probed experimentally, and an upper bound
for the correction was derived.
On the theoretical side, recent advances were based either on asymptotic
expansions valid in the particular case of perfect reflectors at zero
temperature~\cite{Teo2011} or on the derivative expansion approach
\cite{Fosco2011,BimonteEPL2012,BimonteAPL2012}. The latter relies on a
re-summation of the perturbative expansion around the planar geometry. Its
application to compact objects like the sphere thus relies on the assumption
that only the lower hemisphere contributes when $R/L\gg 1$ \cite{Fosco2014}.
Moreover, the derivative expansion requires analyticity of the perturbative
kernel, a condition not met for the zero-frequency contribution when taking the
plasma model~\cite{Mazzitelli2015}.

None of these approaches allow for a direct comparison with the experimental
bound derived in Ref.~\onlinecite{Krause2007}, since they provide only the
leading-order correction to the PFA result. In fact, the next-to-leading-order
correction might be comparable to the leading-order one for typical experimental
aspect ratios $R/L\sim 10^3,$ as for instance in the case of Drude metals at
high temperatures \cite{Bimonte2012}. A recent proposal combines the
leading-order correction for the positive Matsubara frequencies and the exact
result for the zero frequency contribution in the case of Drude metals
\cite{Bimonte2017B}.  However, no such result is available for the
plasma model. Here, we present exact numerical results for the force and the
force gradient taking the parameters of \cite{Krause2007} and either the Drude
or plasma prescriptions for the Matsubara zero-frequency contribution.
Both prescriptions violate the experimental bound for the correction to the
force gradient at sphere-plane distances below 400\,nm, but the corrections for
the Drude prescription are significantly smaller.

In the scattering approach the Casimir interaction free energy is given by
\cite{Emig2007,Lambrecht2006}
\begin{equation}
\label{eq:casimir_F}
\mathcal{F}=\frac{k_B T}{2} \sum_{n=-\infty}^\infty
            \log\det\left[1- \mathcal{M} (\left|\xi_n\right|)\right]\,,
\end{equation}
where $\mathcal{M}(\xi)$ denotes the round-trip operator at imaginary frequency
$\xi$ and the Matsubara frequencies $\xi_n=2\pi n k_B T/\hbar$ are proportional
to the temperature $T$. For reasons explained below and in contrast to the
common choice, we adopt a symmetrized form of the round-trip operator
\begin{equation}
\label{eq:M}
\mathcal{M}(\xi) = \sqrt{\mathcal{R}_S} e^{-\mathcal{K}(L+R)}
    \mathcal{R}_P e^{-\mathcal{K}(L+R)} \sqrt{\mathcal{R}_S} \,.
\end{equation}
The reflection operator at the plane $\mathcal{R}_P$ is diagonal in the
plane-wave basis. Its matrix elements are given by the Fresnel coefficients
$r_p(k, i\xi)$, where the polarization $p$ can either be transverse magnetic
(TM) or transverse electric (TE), and $k$ denotes the projection of the wave
vector onto the plane. The translation operator
$\exp\left[-\mathcal{K}(L+R)\right]$ covers the distance
between the plane and the center of the sphere
along the $z$-direction, cf.\ Fig.~\ref{fig:geometry}a. $\mathcal{K}$ is diagonal in the plane-wave
basis as well, with matrix elements $\sqrt{k^2+\xi^2/c^2}.$ Finally, the reflection
operator $\mathcal{R}_S$ is diagonal in the multipole basis with matrix
elements given by the Mie coefficients $a_\ell(i\xi), b_\ell(i\xi)$
\cite{BohrenHuffman}.

Our particular choice (\ref{eq:M}) for the round-trip operator $\mathcal{M}$ is
a key ingredient to push the numerics into the experimentally accessible
parameter range. First, it avoids ill-conditioned matrices with elements
differing by more than hundred orders of magnitude that render a fast and
stable evaluation of the free energy (\ref{eq:casimir_F}) difficult
\cite{CanaguierThesis}. In fact, numerical tests suggest that with the
round-trip operator of the form (\ref{eq:M}), $1-\mathcal{M}$ becomes
diagonally dominant. Second, it turns out that the matrix $\mathcal{M}$ can be
hierarchically factored. This means that although the matrix is not sparse, it
can be efficiently approximated by considering only a subset of all matrix
elements. The error caused by this approximation can be made negligibly small.
We efficiently compute the determinants using the implementation
\cite{hodlrcode} of an algorithm designed for hierarchical off-diagonal
low-rank matrices \cite{ambikasaran2013}. Another key ingredient is a fast
computation of Legendre polynomials $P_\ell(z)$ \cite{bogaert2012} to
efficiently evaluate associated Legendre functions $P_\ell^m(z)$ arising in the
change between multipole and plane-wave basis. These numerical improvements
allow us to calculate the plane-sphere Casimir energy up to aspect ratios
$R/L\sim 4\cdot10^3$ requiring multipole orders $\ell\sim 2\cdot10^4$.

While the details of our numerical approach will be discussed elsewhere
\cite{HartmannNumerics}, it is worth pointing out checks supporting the validity
of our results. We have found agreement with the exact analytical result for the
Drude model in the high-temperature limit \cite{Bimonte2012} and the leading
correction to PFA for perfect reflectors at $T=0$ \cite{Teo2011,BimonteEPL2012}.
Finally, for the Drude prescription our results shown below in
Fig.~\ref{fig:corrections}b are consistent with those obtained from the
derivative expansion \cite{BimonteAPL2012}.

Here, we will focus on gold surfaces \cite{Krause2007,Decca2007,Decca2007EPJC,Chang2012} at
room temperature $T=295\,\mathrm{K}$. The permittivity of gold at imaginary
frequencies entering the reflection coefficients can be derived from tabulated
optical data \cite{Palik} as explained in Ref.~\onlinecite{Lambrecht2000}. As
shown in Fig.~\ref{fig:geometry}b, the frequency range covered by this
procedure includes all required Matsubara frequencies except for $n=0$.

For the treatment of the zero-frequency contribution, two models have been used
in the analysis of experiments, the Drude model and the plasma model. Since for
$n=0$ no polarization mixing occurs \cite{Umrath2015}, TM and TE modes
contribute independently. The TM mode for both models is
perfectly reflected by plane and sphere, and thus its contribution to the
Casimir free energy only depends on $R/L$. In contrast, the contribution for
the TE mode depends on the model chosen. While for the Drude model no
contribution arises \cite{Bostrom2000}, the contribution for the plasma model
is non-vanishing and also depends on the plasma frequency $\omega_P$
\cite{Canaguier2012}.

It has been argued that even in the plasma model the TE mode does not
contribute to the Casimir free energy \cite{Guerout2014}. Nevertheless, for a
number of experiments agreement of the results with the plasma prescription
just introduced was found
\cite{Decca2007,Decca2007EPJC,Chang2012,Banishev2013,Bimonte2016}.
In few cases experimental support for the Drude prescription was claimed
\cite{Sushkov2011,GarciaSanchez2012}, but also questioned
\cite{BordagComment2012}. All in all, there is no agreement yet on how the
zero-frequency contribution should be accounted for.

For gold, we find from the optical data the plasma frequency
$\omega_P=9\,\mathrm{eV}$ which differs slightly from the value
$\omega_P=8.9\,\mathrm{eV}$ used to analyze the experiment in Ref.\
\onlinecite{Decca2007}. For the zero-frequency contribution in the Drude case, we use
the analytical result derived with the help of bispherical
coordinates~\cite{Bimonte2012} instead of performing a numerical evaluation.

\begin{figure}
 \begin{center}
  \includegraphics[width=\columnwidth]{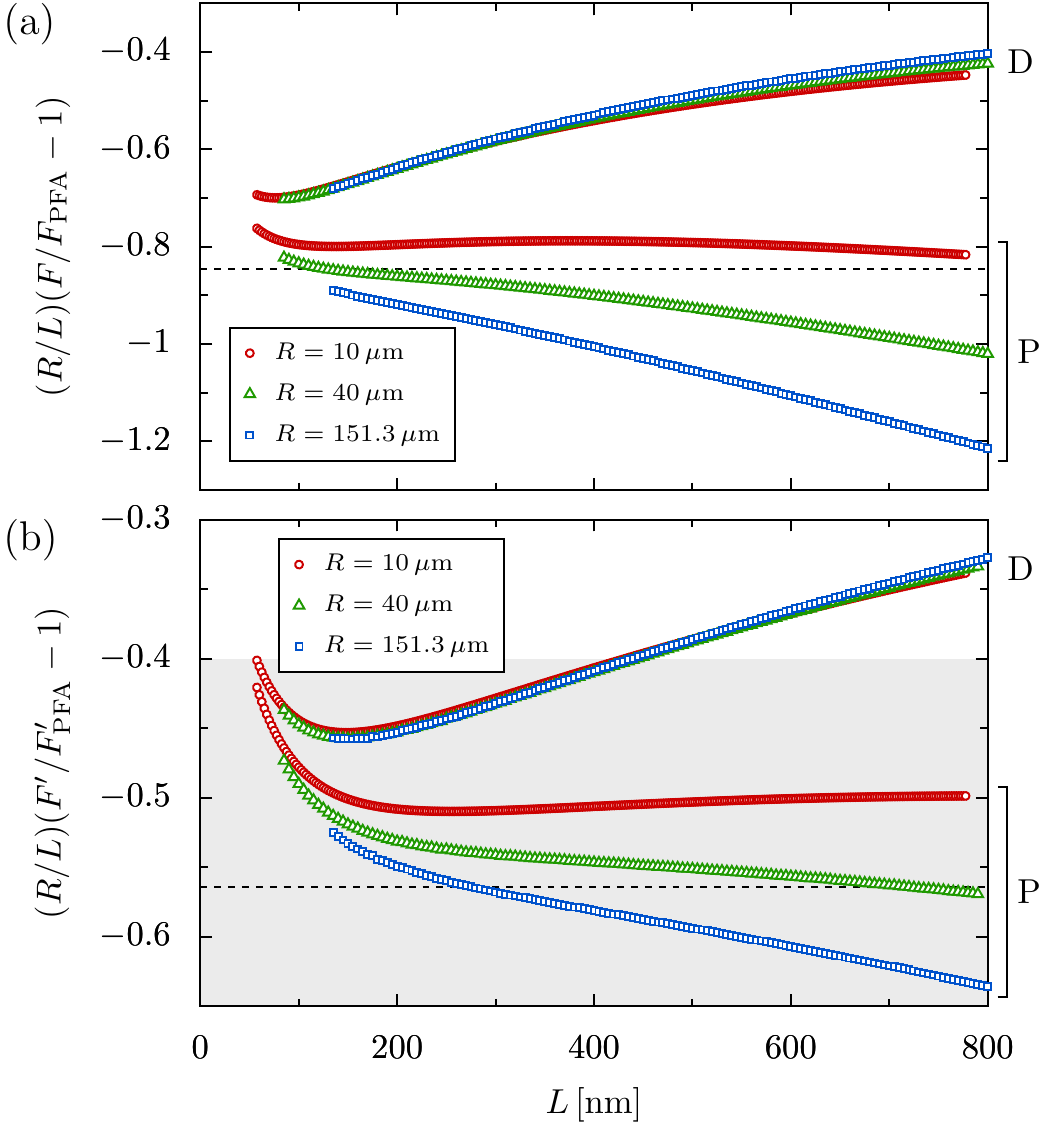}
 \end{center}
 \caption{Beyond-PFA corrections to (a) the force and (b) the force gradient
   are shown as function of the distance between sphere and plane. We multiply
   the correction by the aspect ratio $R/L.$ The upper three and lower three
   lines refer to the Drude (D) and plasma (P) prescription, respectively. The dashed
   horizontal lines indicate the results for perfect reflectors at zero
   temperature, corresponding to the constant coefficients
   $\beta_{T=0}^\text{PR}$ and ${\beta'}_{T=0}^\text{PR}$ for the force and
   force gradient, respectively, as defined in the text. In the lower panel,
   the grey area marks the parameter range for the force gradient excluded
   experimentally at the 95\% confidence level \cite{Krause2007}.}
 \label{fig:corrections}
\end{figure}

We calculate the Casimir force $F = -\partial\mathcal{F}/\partial L$ and compare
the results with the proximity force approximation
$F_{\rm PFA}= 2\pi R \mathcal{F}_{\rm PP}(L)/A$,
where $\mathcal{F}_{\rm PP}(L)/A$ is the Casimir free energy per unit area for parallel plates at a distance $L.$
Within the derivative expansion approach,
the leading correction to PFA is of the form
\begin{equation}
\label{eq:DE}
\frac{F}{F_{\rm PFA}}-1= \beta(L)\frac{L}{R}+\ldots\, ,
\end{equation}
with the coefficient $\beta(L)$ independent of $R$ \cite{Fosco2011}. The
sub-leading corrections might contain logarithmic terms, as for instance
in the case of high temperatures \cite{Bimonte2012}. For the special case
of perfect reflectors and zero temperature, the coefficient $\beta$
is independent of $L$ and given by
$\beta_{T=0}^\text{PR}=1/6-10/\pi^2\approx
-0.847$ \cite{Teo2011,BimonteEPL2012}.

In order to test (\ref{eq:DE}) and obtain a numerical approximation for
$\beta(L),$ we plot in Fig.~\ref{fig:corrections}a the variation of $(R/L)({F}/{F_{\rm PFA}}-1)$ with the
distance $L$ for radii $R=10,40$, and $151.3\,\mu\mathrm{m}$.
The upper three curves correspond to the Drude prescription (D) while the lower three
curves correspond to the plasma prescription (P). The dashed line indicates the
value of $\beta_{T=0}^\text{PR}$. According to (\ref{eq:DE}), the correction
to the force scaled with $R/L$ should approach $\beta(L)$ and be independent of
$R$ for sufficiently small values of $L/R$. This is indeed the case when
considering the Drude prescription for $L\lesssim400$\,nm and the sphere radii shown
in Fig.~\ref{fig:corrections}. As the distance increases, the curves
representing different radii start to deviate from each other. This behavior can be
associated with the contribution of sub-leading corrections, which become
comparatively more important as $L$ increases. At a fixed temperature, larger
distances result in an increase of the relative contribution of the zero
Matsubara frequency \cite{Bimonte2012}, for which the sub-leading correction is
comparable to the leading one for the parameters represented in the figure.

On the other hand, when taking the plasma prescription for the
zero-frequency contribution, the curves corresponding to different values of
$R$, shown in the lower part of Fig.~\ref{fig:corrections}a, are well separated
from each other, indicating that the correction to PFA is not of the form
(\ref{eq:DE}) in this case. The contributions from the Matsubara frequencies
with $n\ne0$ are exactly the same for the two models. Hence the difference
shown in Fig.~\ref{fig:corrections}a is entirely due to the TE zero-frequency
contribution present in the plasma prescription but not in the Drude
prescription.

The zero-frequency contribution becomes relatively more important as $L$
increases, separating the plasma curves from each other and from the Drude
curve. The derivative expansion approach fails in the plasma model at finite
temperatures precisely because of the non-analytical nature of the perturbative
kernel corresponding to the TE zero-frequency
contribution~\cite{Mazzitelli2015}, thus resulting in the structure shown in
the lower part of Fig.~\ref{fig:corrections}a. We also remark that in contrast
to what is frequently believed, the case of perfect reflectors at zero
temperature, indicated by the horizontal dashed line in
Fig.~\ref{fig:corrections}a, does not provide an upper bound for the magnitude
of the force correction for $L\gtrsim100\,$nm due to the contribution of the TE
zero-frequency mode in the plasma model.

The magnitude of the correction to PFA was experimentally investigated in
Ref.~\onlinecite{Krause2007}. The Casimir force gradient $F'=-\partial^2
\mathcal{F}/\partial L^2$ was measured for different sphere radii, and a linear
dependence with $1/R$ similar to (\ref{eq:DE}) was proposed
\begin{equation}
\frac{F'}{F'_{\rm PFA}}-1= \beta'(L) \; \frac{L}{R} + \dots \,.
\end{equation}
While the authors of \cite{Krause2007} were unable to measure the correction term, they
nonetheless derived the upper bound $|\beta'(L)|<0.4$ at the 95\% confidence
level for $L$ in the interval between $150$ and $300\,\text{nm}$. In
Fig.~\ref{fig:corrections}b, we plot the variation of $(R/L)({F'}/{F'_{\rm
PFA}}-1)$ with the distance $L$ for the same values of $R$ used in
Fig.~\ref{fig:corrections}a. This quantity provides an approximation for the
coefficient $\beta'(L)$ as long as the results are independent of $R.$ The
shaded area represents values for the correction excluded by the experiment
\cite{Krause2007} while the dashed line indicates the correction for perfect
reflectors at $T=0$ given by ${\beta'}_{T=0}^\text{PR}=(2/3)\beta_{T=0}^\text{PR}\approx-0.564$ \cite{Teo2011,BimonteEPL2012}.

For $L\lesssim 400$\,nm the Drude as well as the plasma prescription violate the
experimental bound, although the maximum violation for the Drude prescription at
$L\approx150\,\text{nm}$ corresponding to $\beta'\approx -0.46$ is significantly
smaller than the violation found for the plasma prescription. Note, however, that the
plasma and Drude curves get closer to each other as the distance decreases below
$200\,\text{nm}$, as expected in the low temperature regime, with the zero
frequency providing a relatively smaller contribution.

As in the discussion of the correction to the force, the results for different
radii shown in Fig.~\ref{fig:corrections}b are very close to each other and to
the results obtained within the derivative expansion approach
\cite{BimonteAPL2012, Bimonte2017B} when taking the Drude prescription. In this case, our results
show that sub-leading corrections are negligible for the experimental
conditions of Refs.~\onlinecite{Krause2007,Decca2007,Decca2007EPJC,Chang2012}, which correspond to
aspect ratios in the range $R/L\sim 10^2-10^3.$ As a consequence, the
corrections can be directly obtained within the derivative expansion approach
\cite{BimonteAPL2012,Bimonte2017B}. However, for the plasma prescription the derivative expansion
clearly underestimates the correction, particularly for the largest radius
shown in Fig.~\ref{fig:corrections}b, and the leading order correction is not
proportional to $1/R.$

\begin{figure}
 \begin{center}
  \includegraphics[width=\columnwidth]{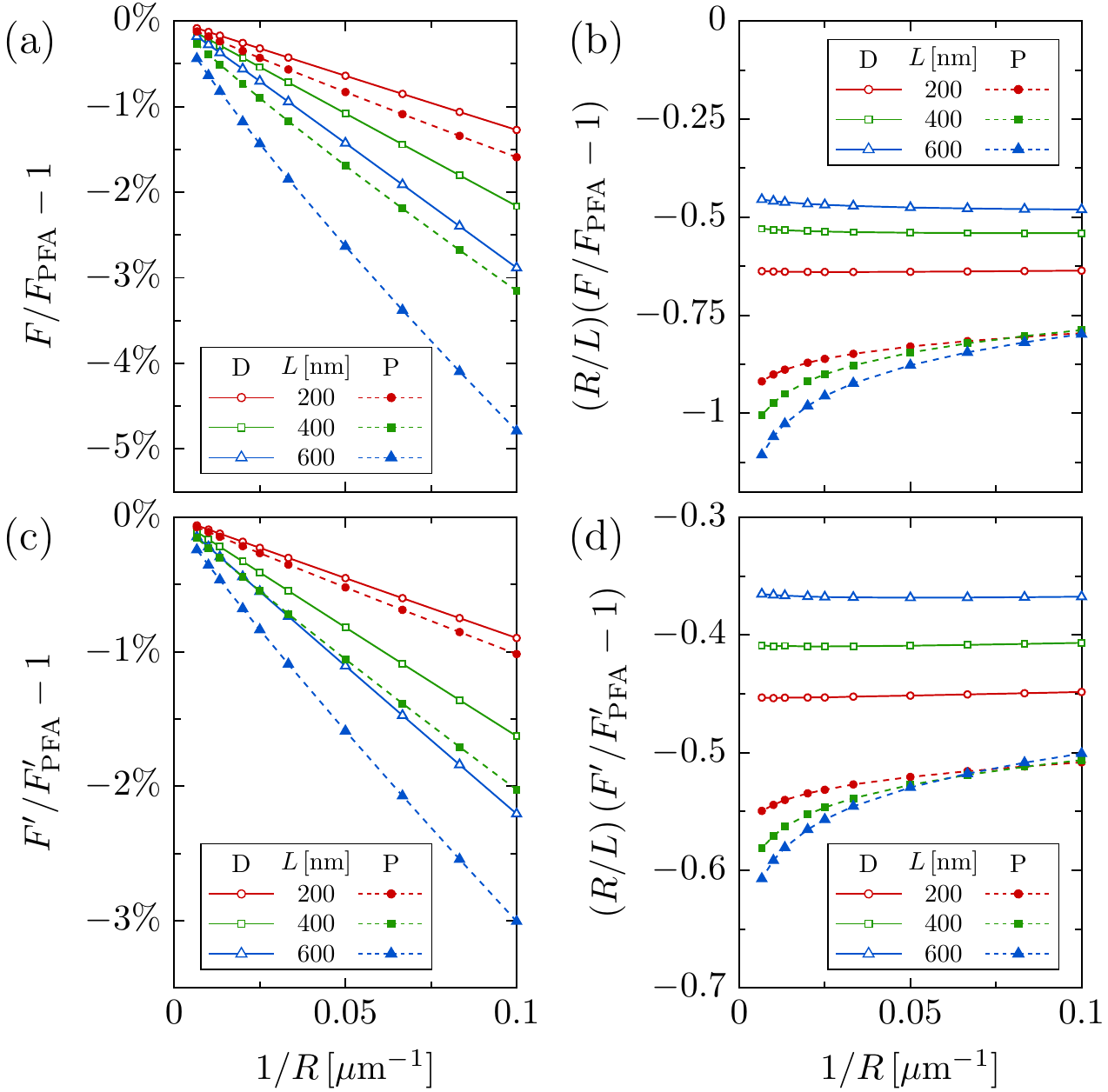}
 \end{center}
 \caption{Beyond-PFA corrections to the (a,b) force and (c,d) force gradient
   are shown as function of the inverse sphere radius. While in panels (a) and
   (c) the relative corrections are displayed, the data in panels (b) and (d)
   have been scaled by $R/L.$ The sphere-plane distances are $L=200\,\text{nm}$
   ($\circ$), $400\,\text{nm}$ ($\square$), and $600\,\text{nm}$ ($\triangle$).
   Solid lines with open symbols refer to the Drude prescription while dashed lines
   with filled symbols refer to the plasma prescription.}
 \label{fig:plasma_r2}
\end{figure}

In order to better understand the dependence on the sphere radius, we plot in
Figs.~\ref{fig:plasma_r2}a and \ref{fig:plasma_r2}c the force and force gradient
corrections, respectively, as function of $1/R$. For the plasma prescription,
the force corrections are typically close to or above the percent level for the
conditions of the experiment \cite{Chang2012} where $1/R=0.0242\,\mu$m$^{-1}$.
More importantly for this experiment, the corrections to the force gradient are
typically below $1\%$ for sub-micrometer distances. In
Figs.~\ref{fig:plasma_r2}b and \ref{fig:plasma_r2}d the corrections to the force
and force gradient, respectively, are scaled by $R/L$. While for the Drude
prescription the data follow rather closely a $1/R$ dependence, the results for
the plasma prescription indicate a more singular approach to the PFA limit as
$1/R\rightarrow 0$.

In conclusion, we have shown that the Drude prescription for the
Matsubara zero-frequency contribution leads to a weaker violation of the upper
bound for the PFA correction derived experimentally by measuring the force
gradient for different radii \cite{Krause2007} than the dissipationless plasma
prescription. This could have been expected, since dissipation is present
in the gold coatings used in the experiments. However, all experiments
performed with coated microspheres with aspect ratios $R/L\sim 10^2 - 10^3$
agree with the plasma prescription but not with the Drude prescription when the force
variation with distance is analyzed for a given radius
\cite{Decca2007,Decca2007EPJC,Chang2012,Banishev2013,Bimonte2016}. The proximity force approximation
combined with the Drude prescription underestimates the experimental data for
nonmagnetic materials, so that the correction calculated here brings the Drude
prediction even further away from the experimental results. When taking the
plasma prescription, the magnitude of the correction is significantly larger than
predicted experimentally but still too small to degrade the quality of the
comparison between the experimental data and the theory based on the plasma
prescription.
The theoretical results presented here, taking the sphere curvature fully into
account, indicate that experiments probing the Casimir interaction beyond the
PFA regime could provide new insight into the role of dissipation in Casimir
physics.

We thank A. Canaguier-Durand, R. Gu\'erout, A. Lambrecht, S. Reynaud, G.
Bimonte, and F. D. Mazzitelli for discussions, and D. Dalvit for providing
numerical data for the permittivity of gold along the imaginary frequency axis.
We acknowledge support from CAPES and DAAD through the PROBRAL collaboration
program. PAMN also thanks CNPq and FAPERJ for partial financial support.

\end{document}